\documentclass[aps,twocolumn,showpacs]{revtex4}
\usepackage{graphicx}
\begin{document}

\title{Electronic transport in an anisotropic Sierpinski gasket}

\author{Supriya Jana$^1$, Arunava Chakrabarti$^1$ and Samar Chattopadhyay$^2$}

\affiliation{$^1$Department of Physics, University of Kalyani, Kalyani, 
West Bengal-741 235, India. \\
$^2$Department of Physics, Maulana Azad College \\ 8, Rafi Ahmed Kidwai Road, 
Kolkata - 700 013, India.}

\begin{abstract}
We present exact results on certain electronic properties of an
anisotropic Sierpinski 
gasket fractal. We use a tight binding Hamiltonian and work within the 
formalism of a real space renormalization group (RSRG) method. The anisotropy is 
introduced in the values of the nearest neighbor hopping integrals.  
An extensive numerical examination of the two terminal transmission spectrum and 
the flow of the hopping integrals under the RSRG iterations strongly suggest that 
an anisotropic gasket is more conducting than its isotropic counter part and that, 
even a minimal anisotropy in the hopping integrals generate {\it continuous bands} of 
eigenstates in the spectrum for finite Sierpinski gaskets of arbitrarily large size.
We also discuss the effect of a magnetic field threading the planar gasket on its 
transport properties and calculate the persistent current in the system. The sensitivity 
of the persistent current on the anisotropy and on the band filling is also 
discussed. 
\end{abstract}

\pacs{73.21.-b, 73.22.Dj, 73.23.Ad, 73.23.Ra}

\maketitle 

\noindent
\section{INTRODUCTION}

Electronic properties of fractal lattices have been extensively investigated 
in the past~\cite{domany}-\cite{meyer}, and a wealth of knowledge has been 
accumulated. Such deterministic lattices provide excellent examples of 
systems in between perfect periodic order and complete randomness and exhibit 
electronic properties drastically different from a crystal and a disordered 
material. The variable environment around any site of a 
fractal structure makes such systems quite similar to the random networks. This 
has motivated a series of experiments to examine the fractal behavior, magneto-resistance 
and the superconductor-normal phase boundaries on Sierpinski gasket wire 
networks~\cite{gordon1,gordon2,gordon3,korshu,meyer}. 
The interface of theory and experiments has further been strengthened by 
the recent development of non-dendritic, 
perfectly self-similar fractal macromolecules of hexagonal Sierpinski 
gaskets~\cite{new}.

Fractals, in general, exhibit a Cantor set like energy spectrum 
~\cite{domany} that is highly degenerate, and the density of states displays 
a wide variety of singularities. The spectrum may however, broaden up in the 
presence of a magnetic field~\cite{banavar}. The electronic conductance 
typically exhibits scaling with a multifractal distribution of the exponents
~\cite{schwalm2}. One useful tool in investigating 
regular, finitely ramified fractal lattices has been the real space renormalization 
group (RSRG) methods.  
The RSRG approach has been successfully used over the years to unravel, 
for example, 
the spectral features of Koch fractals 
with long range interactions~\cite{maritan}, localization aspects and the length 
scaling of corner-to-corner propagation in fractal glass and several other interesting 
fractal networks~\cite{schwalm1,andrade3}.

Interestingly, RSRG studies on fractal lattices reveal another remarkable 
property of such systems, viz, the existence of an infinite number of 
isolated {\it extended} eigenstates that coexist with an otherwise Cantor set 
spectrum. This is non trivial, as the fractals as such do not have any long range 
translational order. Such extended states sometimes are traced back to the 
existence of multiple cycle fixed points of the Hamiltonian, 
or have been shown to arise out of a local positional correlation, giving rise to 
a resonant tunneling effect in some local atomic clusters~\cite{wang2,arun1,arun2,arun3}. However, 
no conclusive evidence of the possible formation of a continuum of states has been 
obtained until very recently, when Schwalm and Moritz~\cite{schwalm4} 
have presented exhaustive numerical results to show that a family of fractals 
does possess a continuum of such states. Incidentally, such continua were also 
`suspected' in some of the previous studies~\cite{arun1,arun2,arun3}, where an 
external magnetic field was shown to play a role.

\begin{figure}[ht]
{\centering\resizebox*{8cm}{5cm}{\includegraphics{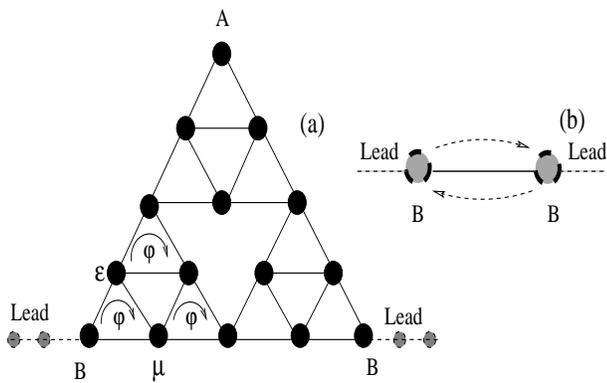}} \par}
\caption{A Sierpinski gasket and its renormalization to a `dimer'}
\label{gasket}
\end{figure}

This, to our mind, keeps the problem alive, and we look back into the
aspect of observing continuous distribution of eigenstates in an anisotropic  
Sierpinski gasket (SPG) fractal. The magnetic flux, as discussed in the earlier works 
\cite{arun1,arun2,arun3} through 
each elementary plaquette of an isotropic SPG breaks the time reversal symmetry of the 
electron hopping between nearest neighbors.
This in turn, already generates a kind of anisotropy in the hopping of 
an electron when it travels along an {\it angular} bond, compared to 
when it hops along a {\it horizontal} bond (Fig.~\ref{gasket}). Is 
anisotropy the factor responsible for any continuum then ? 
This question is the central motivation behind our present work.
Such an issue 
had been addressed earlier by Hood and Southern~\cite{hood} using a 
generating function approach coupled with RSRG to come to the conclusion that 
in an anisotropic SPG only certain {\it hierarchical} states persist.

A Second motivation is of course the fact that, grafting a deterministic fractal 
geometry on a given substrate
is quite feasible using the present day advanced nanotechnology and
lithographic methods. Therefore, any spectral peculiarity that might come out
as a consequence of the anisotropy, and is likely to get reflected in the
transport behavior, would be quite possible to observe. This may throw some
light into the unaddressed issues related to such deterministic fractal lattices, and
will definitely enrich the experimental aspects.

In this communication We examine the transmission spectrum of 
finite but arbitrarily large anisotropic SPG lattices, with and 
without any magnetic flux threading the system, using an RSRG decimation 
method~\cite{arun1}. In addition to this, we also calculate the persistent current in 
such a network, and look for any peculiarities in the behavior of the 
current as well as in the Aharonov-Bohm (AB) oscillations in the transmission 
that might come out as a result of the anisotropy. To the 
best of our knowledge, 
no results are available for persistent currents in an SPG network, 
specially for the anisotropic case, and the impact of the scale invariant, 
multiple loop geometry of an SPG on such properties 
are either very little addressed, or unaddressed at all. 
Therefore, a thorough study of the problem, together with the literature that already 
exists will possibly help in obtaining some conclusive result about the spectra 
of regular fractal lattices under various conditions.

We find interesting results. Anisotropy indeed signals the 
formation of {\it bands} of eigenstates with high transmittivity. 
In general, it is seen that an anisotropic SPG is more conducting compared to its 
isotropic counterpart. We have carefully carried out an RSRG analysis of the 
recursion relations of the parameters of the Hamiltonian, and observe certain 
unusual flow pattern. The transmission coefficient for arbitrarily large 
lattices has been worked out. In addition to this, we present an in-depth 
study of the persistent current observed in a SPG network, and the influence of the 
degree of anisotropy on the current. The behavior of the persistent current is 
seen to be very sensitive to the degree of anisotropy. 

In what follows, we present the results of our investigation. In section II, we present 
the model and the RSRG method. Section II contains the discussion on the transmission 
spectrum, while, in section IV we present the results of the calculation of the persistent 
current. 
\vskip .3in
\section{THE MODEL AND THE METHOD}

We begin by referring to Fig.~\ref{gasket}. The Hamiltonian of the  
network is given by, 
\begin{equation}
{\mathbf H}=\sum_n \epsilon_n c_n^{\dagger} c_n + t e^{i\theta} \sum_{<nm>} c_n^{\dagger} 
c_{m} + h.c. 
\label{equ1}
\end{equation}
where, 
$c_{n}$ ($c_{n^{\dagger}}$) is the annihilation (creation)
operator at the $n$th site of the gasket, the nearest neighbor hopping 
integral $t = t_x$ along the `horizontal' bond and it is equal to $t_y$ along any 
`angular' bond. $\theta = 2 \pi \phi/\phi_0$ is the phase acquired by the 
electron in hopping along a side of the elementary triangle, as shown in Fig.~\ref{gasket}.
The phase factor $\theta$ is positive if the electron hops around an elementary 
triangle in the clockwise sense, and negative otherwise. 
We have chosen two kinds of `on-site' potentials depending on the local environment of any 
given site. They are symbolized as $\epsilon$ and $\mu$, and the `edge' sites have 
potentials $\epsilon_A$ and $\epsilon_B$ corresponding to the vertices $A$ and $B$.  
For calculating the end-to-end transmission we have fixed two 
semi-infinite ordered chains as leads (Fig.1) at the end sites $B$. The on-site 
potential at the lead-sites has been taken to be constant and equal to $\epsilon_0$. 
The nearest neighbor hopping integral in the leads 
is chosen as $t_0$.

To obtain the transmission coefficient, as well as to judge the nature of the 
eigenfunctions we renormalize the SPG using a decimation 
technique~\cite{arun1}. We first explicitly discuss the anisotropic SPG 
{\it without any magnetic field}. 
The recursion relations for the on-site potentials and 
the hopping integrals ( in the absence of any flux threading the gasket) 
are given by, 
\begin{widetext}
\begin{eqnarray}
\epsilon_{n+1} & = & \epsilon_n + \frac{p_n + 2q_n + r_n}{\alpha_n \beta_n -
\gamma_n} + s_n \nonumber \\
\mu_{n+1} & = & \mu_n + u_n + v_n \nonumber \\
\epsilon_{A,n+1} & = & \epsilon_{A,n} + \frac{(E - \epsilon_n) [ \beta_n t_{y,n}^2 
+ \alpha_n t_{x,n}^2 + 2 t_{x,n}^2 t_{y,n}^2 (E - \epsilon_n + t_{x,n}) ]}
{\alpha_n \beta_n - \gamma_n} \nonumber \\
\epsilon_{B,n+1} & = & \epsilon_{B,n} + \frac{\beta_n t_{y,n} (E - \epsilon_n + t_{x,n}) + 
\gamma_n t_{y,n}}{\alpha_n \beta_n - \gamma_n} \nonumber \\
t_{x,n+1} & = & \frac{[ (E - \epsilon_n) t_{x,n} + t_{y,n}^2]^2 + t_{x,n} t_{y,n}^2 (E - \mu_n + 2t_{x,n}) 
- t_{x,n}^4}{( E - \epsilon_n + t_{x,n} ) [ \beta_n - t_{y,n}^2 - t_{x,n} ( E - \mu_n) ]} \nonumber \\
t_{y,n+1} & = & \frac{t_{y,n}^2 ( E - \mu_n + 2 t_{x,n} )}{\beta_n - t_{y,n}^2 - t_{x,n} ( E - \mu_n)}
\end{eqnarray}
\end{widetext}
Here, we have defined,
\begin{eqnarray}
\alpha_n & = & (E - \epsilon_n)^2 - t_{x,n}^2 \nonumber \\ 
\beta_n & = & (E - \epsilon_n) (E - \mu_n) - t_{y,n}^2 \nonumber \\
\gamma_n & = & t_{y,n}^2 (E - \epsilon_n + t_{x,n})^2 \nonumber \\
p_n & = & t_{y,n}^2 \beta_n (E - \epsilon_n) \nonumber \\
q_n & = & t_{x,n} t_{y,n}^2 (E - \epsilon_n) (E - \epsilon_n + t_{x,n}) \nonumber \\ 
r_n & = & t_{x,n}^2 \alpha_n (E - \epsilon_n) \nonumber \\
s_n & = & \frac{2t_{y,n}^2 (E - \epsilon_n)}{\beta_n - t_{y,n}^2 - t_{x,n}(E - \mu_n)} \nonumber \\
u_n & = & \frac{2t_{x,n}^2 (E - \epsilon_n + t_{x,n}) + t_{y,n}^2 (E - \mu_n)}{(E - \mu_n)
(E - \epsilon_n + t_{x,n})} \nonumber \\ 
v_n & = &\frac{t_{y,n}^2 (E - \mu_n + 2t_{x,n})^2}{(E - \mu_n)[\beta_n - t_{y,n}^2 - t_{x,n}(E - \mu_n)]} 
\end{eqnarray}

Using the above set of equations, we renormalize an $n$-th generation SPG $(n-1)$- times 
to reduce it to a simple triangle. The top vertex is then decimated to generate an 
effective diatomic molecule $B-B$ clamped between the leads, as shown in Fig.1(b).
The transmission coefficient is then obtained by the standard formula~\cite{stone},
\begin{widetext}
\begin{equation}
T = \frac{4 \sin^2(qa)}{[M_{1,2} - M_{2,1} + ( M_{11} - M_{22} ) \cos (qa)]^2 + 
(M_{11} + M_{22})^2 \sin^2(qa)}
\end{equation}
\end{widetext}
where, the matrix elements corresponding to the `diatomic molecule' are given by, 
$M_{11} = (E - \epsilon_{B,n})^2/(t_0 t_{x,n}) - t_{x,n}/t_0$, 
$M_{12} = -M_{21} = -(E - \epsilon_{B,n})/t_{x,n}$, and $M_{22} = -t_0/t_{x,n}$.
Before ending this section we just point out that the same decimation technique has been 
used to obtain the recursion relations in the presence of a magnetic field.  
The recursion relations in the latter case are much more complicated compared to the field-free 
case,  
and we decide not to present them here to save space. The role of the magnetic field will  however, 
be discussed in appropriate places.
The results are now presented below.

\begin{figure}[ht]
{\centering\resizebox*{15cm}{12cm}{\includegraphics[angle=-90] {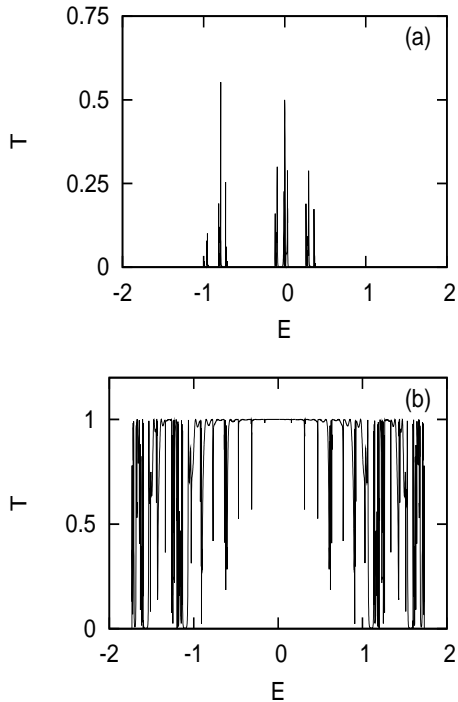}} \par}
\caption{Transmission across a sixth generation 
isotropic gasket for (a) $\phi = 0$, 
and for (b) $\phi = \phi_0/4$. We have chosen $\epsilon_j = 0$ for all 
the vertices of the SPG as well as for the sites in the leads, and 
$t_x = t_y = t_0 = 1$ throughout.}
\label{trans1}
\end{figure}

\section{RESULTS AND DISCUSSION}
\subsection{The transmission spectrum and the electron states}

\begin{figure}[ht]
{\centering\resizebox*{15cm}{12cm}{\includegraphics[angle=-90]{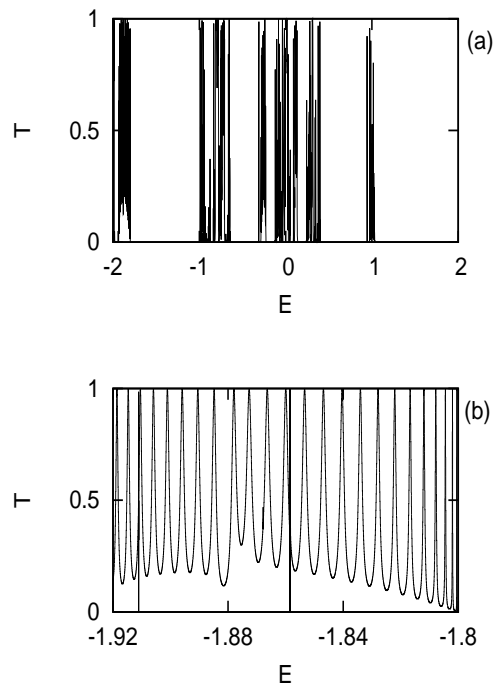}} \par}
\caption{(a) Transmission across a sixth generation 
anisotropic gasket for $\phi = 0$ and with $t_x = 0.9$ and $t_y = 1$.
(b) Fine scan of the energy interval 
$[-1.92,-1.8]$ with $\delta E = 10^-7$. Other parameters are the same as in 
Fig.2, and energy is measured in unit of the hopping in the leads. 
}
\label{trans2}
\end{figure}

The transmission coefficients for the zero-field case and for 
flux $\phi = \phi_0/4$ threading an elementary triangle have been shown in 
Fig.~\ref{trans1} for a sixth generation SPG in the isotropic limit 
for the sake of completeness. As is already 
known~\cite{arun1}, the magnetic field makes an isotropic gasket 
more conducting with the signature of {\it continuous} zones of transmission. 
When a slight anisotropy is introduced, and the magnetic field is taken off, 
it is interesting to observe that, first of all the SPG becomes much more 
conducting than its isotropic counterpart, and second, once again patches of 
continuum appear in the transmission spectrum. One such example is presented 
in Fig.~\ref{trans2} with $t_x = 0.9$ and $t_y = 1$. 
With increasing degree of anisotropy 
the continua become much more pronounced at places. We have not been able to 
provide an analytical proof as to whether real continua appear even in such 
non-translationally invariant systems, but have carried out extensive numerical 
investigations involving very fine scan of the `apparently' continuous bands 
found in the transmission spectrum. In every case we find the fine structure 
in these `special' zones free from any real gaps. 
Even, the apparently sharp peaks in the transmission spectrum show up 
finite width when the energy is scanned minutely in their neighborhood.
Thus, it is tempting to 
claim that anisotropy introduces {\it continuous bands} of high transmission 
in fractals such as an SPG. In a recent work  
Schwalm and Moritz~\cite{schwalm4} have discussed precisely this issue in the 
case of a different class of hierarchical lattices. This latest observation gives us 
confidence in regards of the existence of bands in fractal lattices which can 
coexist with even a fragmented Cantor like spectrum. 

\begin{figure}[ht]
{\centering\resizebox*{8cm}{8cm}{\includegraphics {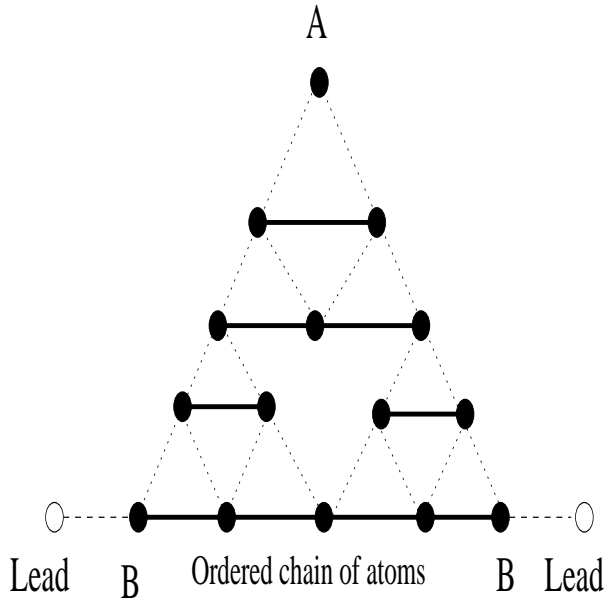}} \par}
\caption{Separation of  a finite Sierpinski gasket under RSRG into 
detached diatomic clusters and linear ordered chain clamped between the leads.} 
\label{rsrg}
\end{figure}

Are these conducting states extended ?     
To understand this, we have critically examined the flow of the hopping 
integrals $t_x$ and $t_y$ under successive RSRG steps. We find two different 
types of behavior. In general, for an arbitrarily chosen energy both 
$t_x$ and $t_y$ flow to zero quickly under iteration. This is understandable, as 
the spectrum of such a fractal lattice is fragmented in general. So, any 
arbitrarily chosen energy, in all probability, would either fall in a gap, or 
would correspond to a localized eigenstate. 
However, Hood and Southern~\cite{hood} have eliminated the possibility of 
exponentially localized states in an anisotropic SPG.
The flow of the hopping integrals under RSRG changes its pattern when 
we choose an energy from what apparently looks like a continuum in the transmission 
spectrum in Fig.~\ref{trans2}(a). For any energy within the {\it continuous} portion 
of the spectrum $t_x$ remains non-zero for arbitrary number of 
RSRG loops while the hopping $t_y$ ultimately flows to zero. 
This feature is always true no matter whether the initial 
value of $t_x$ is less or greater than the initial value of $t_y$, and persists 
for any degree of anisotropy. That is, the anisotropy somehow develops a `preference' 
for $t_x$, and breaks the symmetry at all scales of length. 
Thus, in the infinite lattice limit the entire SPG 
fractal consists of isolated {\it dimers} coupled by the hopping integral 
$t_{x,n \rightarrow \infty}$. For a finite (but arbitrarily large) SPG we have 
coupled the semi-infinite leads to the two base atoms ($B-B$). Thus, as $t_y$ becomes 
(practically) zero under RSRG operation, we are left with a perfectly ordered $1$-d 
chain of atoms with an effective, energy dependent  potential connected to its 
nearest neighbor by a hopping amplitude $t_{x,n \rightarrow \infty}$ (Fig.~\ref{rsrg}). 
So, any energy $E$ that will map the original 
SPG to such a configuration will definitely correspond to a conducting state provided, 
the energy falls within the allowed band of the leads. What is peculiar about such energy 
values is that, they apparently form a completely continuous band. We have scanned the 
spectrum very carefully. 
One such result of scanning is exhibited in Fig.~\ref{trans2}(b), where 
we have chosen, quite arbitrarily, a certain portion of the spectrum shown in Fig.2a, and 
scanned the selected zone in an energy interval of $10^{-7}$. It is obvious that the 
spectrum retains its continuous character.

Before we end this sub-section, it is worth mentioning that, getting a non-zero transmission 
is definitely connected to the fact that we have attached the leads at the two base atoms 
of the SPG. Talking about an infinite SPG, we should appreciate that we can build an infinite 
SPG using a {\it top down} approach, when one takes a triangle and keeps on 
piercing it in the SPG design an infinite number of times. We can keep the leads attached to 
the base pair of $B$ atoms from the beginning. So, the effective ordered chain of atoms 
which the SPG boils down to, when we pick up an energy from the continuum, will always 
be found clamped between the leads. The corresponding state will be extended in the sense that 
a finite number of sites of the parent SPG will have non-zero amplitudes of the wave function.
In the reverse approach, when one builds an infinite SPG by placing SPG's of previous generations 
on top of each other, following the growth rule, it is of course not very meaningful to talk 
about the base atoms such as the $B-B$ pair, or even fixing the leads at the ends. So, at most the 
infinite system can break up into diatomic clusters as depicted in Fig.~\ref{rsrg}. 

\subsection{The Aharonov-Bohm oscillations}

\begin{figure}[ht]
{\centering\resizebox*{15cm}{12cm}{\includegraphics [angle=-90]{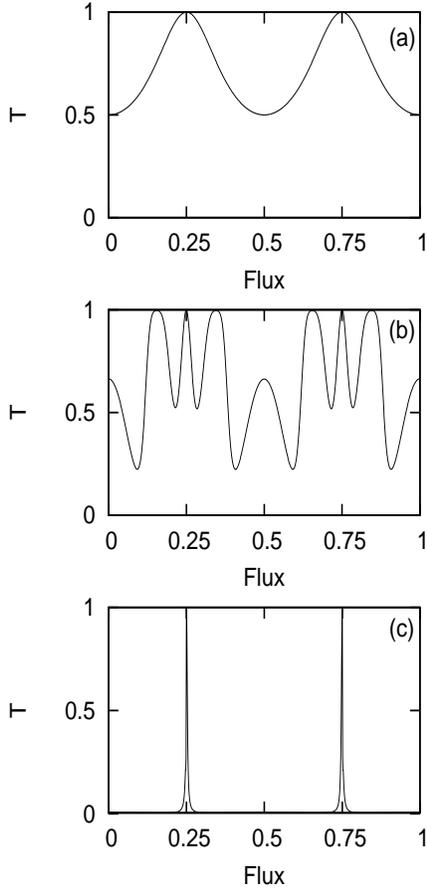}} \par}
\caption{(a) Aharonov-Bohm oscillations in a $4$th generation SPG for $E = 0$.
(a) Isotropic SPG with $t_x = 1$ and $t_y = 1$, (b) anisotropic SPG with 
$t_x = 1$, and $t_y = 2$, and (c) anisotropic SPG with $t_x = 1$, and $t_y = 10$.
The potential and other parameters are the same as in the previous figures.}
\label{ab}
\end{figure}

The anisotropic SPG displays a wide variety of AB-oscillations which are 
sensitive to the energy chosen, as well as on the degree of anisotropy. The 
period of oscillations has been found to be equal to $\phi_0/2$ if we choose, 
say, $E = 0$. The AB-oscillations are displayed in Fig.~\ref{ab} for the 
cases $(t_x = t_y =1)$, $(t_x = 1, t_y = 2)$ and $(t_x = 1, t_y = 10)$. 
The simple oscillation profile in the isotropic case develops into a profile 
with multiple peaks and valleys (Fig.~\ref{ab}(b)) as the anisotropy grows 
larger and finally, in the case of very large degree of anisotropy, the AB-oscillation 
profile consists of delta-like peaks at special values of the magnetic flux, indicating 
that, the anisotropic gasket triggers ballistic transmission for $E = 0$ only at certain 
special values of the magnetic flux.

\subsection{The persistent current}

\begin{figure}[ht]
{\centering\resizebox*{7cm}{14cm}{\includegraphics{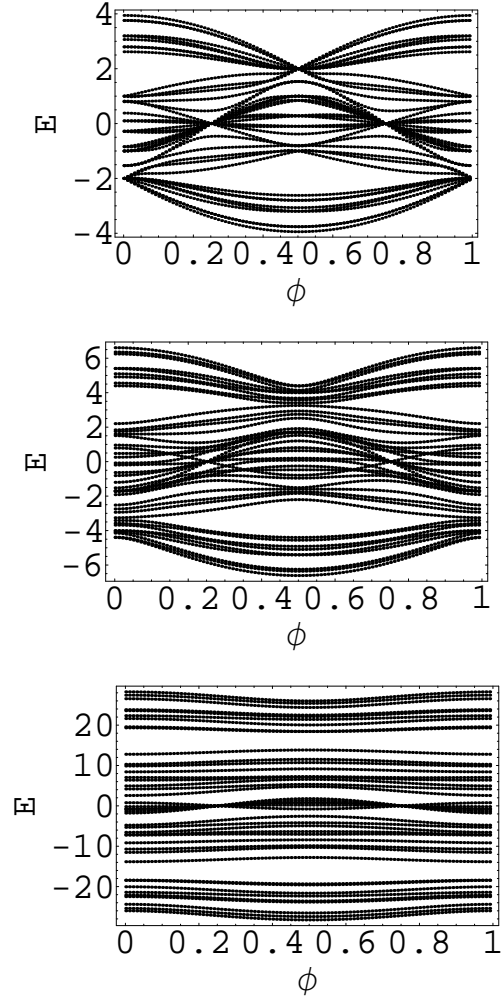}} \par}
\caption{Flux dependence of the eigenvalue spectrum of a $4$th generation
Sierpinski gasket. (a) The isotropic gasket with $t_x = t_y =1$,  
(b) the anisotropic gasket with $t_x =1$, $t_y = 2$, and (c) with 
$t_x =1$ and $t_y = 10$. In each cast the on-site potential is taken 
to be zero throughout.} 
\label{spectrum}
\end{figure}

It is well known that a simple Aharonov-Bohm flux threading a metallic or 
semiconducting ring generates an equilibrium persistent current~\cite{gefen}-
\cite{jin}. With reference to our SPG fractal, where the same flux penetrates 
each elementary triangular plaquette, the eigenvalues and the eigenstates are 
flux periodic with period $\phi_0$,and hence, as is well known, the equilibrium 
persistent current carried by the energy level $E_n$ is given by, 
\begin{equation}
I_n = -c \frac{\partial E_n}{\partial \phi}
\end{equation}
and, the total current in the ring is, $I (\phi) = \sum_{n=1}^{N_e} I_n$ where, 
$N_e$ is the total number of electrons in the system. It is clear that the nature 
of the energy spectrum plays a major role in dictating the persistent current in a 
system. We therefore first have a look at the 
flux dependence of the eigenvalue spectrum of a $4$-th generation 
SPG fractal (Fig.~\ref{spectrum}). For the sake of comparison, we present the 
spectrum of the SPG in the isotropic limit (Fig.~\ref{spectrum}, top panel), which 
shows multiple band crossings, as observed by Rammal and Toulose~\cite{rammal} 
earlier. The band crossings, related to the symmetry and degeneracy of the eigenvalues 
at various values of $\phi/\phi_0$ are already discussed in details in reference~\cite{rammal}.  
\begin{figure}[ht]
{\centering\resizebox*{15cm}{14cm}{\includegraphics{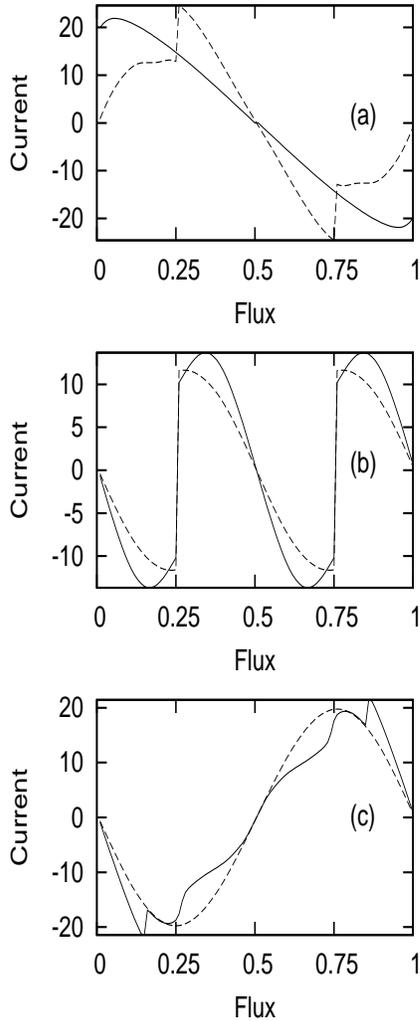}} \par}
\caption{Persistent current in a $4$th generation
Sierpinski gasket. (a) The isotropic gasket with $t_x = t_y =1$. The solid 
and the dashed curves correspond to $N_e = 4$ and $15$ respectively.  
(b) The anisotropic gasket with $t_x =1$, $t_y = 2$ (solid line), and 
$t_x = 1$, $t_y = 10$ (dashed line) with $N_e = 21$, and (c) Anisotropic 
gasket with $N_e = 25$ when  
$t_x =1$, $t_y = 2$ (solid line) and $t_x = 1$, $t_y = 10$ (dashed line). 
In each cast the on-site potential is taken 
to be zero throughout.} 
\label{current}
\end{figure}

With the introduction of anisotropy in the hopping integrals ( $t_x \ne t_y$ ) gaps start 
to open up in the spectrum lifting the degeneracy in places. As the degree of anisotropy is 
increased, the spectrum shows clear signature of having groups of states lying close in energy 
and separated by wider global gaps, that bring a flavor of sub-band structures in the spectrum. 
The middle and the bottom panels of Fig.~\ref{spectrum} represent the anisotropic situation 
with $t_x = 1$, $t_y = 2$  and  $t_x = 1$, $t_y = 10$ respectively, the last case may be taken 
as a case of large anisotropy. The flattening of the bands in this case is noteworthy, which 
finally brings in non trivial changes in the persistent current.

We have examined in details the variation of the persistent current for both the isotropic 
and anisotropic SPG's. for the isotropic case with $t_x = t_y = 1$, the current does not 
exhibit serious deviations (in its qualitative features) from that in a single loop ~\cite{gefen}
as long as $N_e$ is small. With increasing values of $N_e$, a slight rounding off of the 
$I(\phi)$ is observed. The overall current keeps on increasing as $N_e$ increases. The first 
variation shows up for $N_e = 10$, when additional kinks appear at $\phi = \phi_0/2$.
With increasing $N_e$ the SPG starts displaying features different from what one finds in 
simple loop structures. The features persist as we approach the half filled case which, in 
this study is $N_e = 21$, when a precise $\phi_0/2$ periodicity in $I(\phi)$ is observed. 
In Fig.~\ref{current}(a) we show the variation of the persistent current when $N_e=4$ (solid line) 
and $N_e=15$ (dashed line).

Introduction of anisotropy in the values of the hopping integral brings in different 
features in the $I(\phi)$-$\phi$ curves. Though several basic features such as the 
$\phi_0$ periodicity, or the special case of $\phi_0/2$ period at $N_e=21$ persist, the 
degree of anisotropy, i.e. the relative values of $t_x$ and $t_y$ strongly influence the 
current. Needless to say, the filling factor also plays a crucial role in determining the 
current profiles. For example, we have found that with $(t_x,t_y) = (1,10)$, the current drops 
compared to the case when $(t_x,t_y) = (1,2)$ when we take $N_e = 21$. This has been tested 
with several values of the hopping and we come to the conclusion that at half filling, the 
effect of anisotropy is to reduce the overall current. The current profile becomes smooth 
with the discontinuous features of isotropic cases for low $N_e$ as we introduce anisotropy. 
This is mainly due to the gaps which open up in the spectrum as the anisotropy is introduced. 
In this aspect, anisotropy plays a role equivalent to the role that disorder plays in one 
dimensional rings.

\begin{center}
{\bf Acknowledgment}
\end{center}

The authors are thankful to Santanu K. Maiti for helping out with the graphics, and 
useful comments on the results of this paper.

\end{document}